# Graphene Spin Valve Devices

Ernie W. Hill, *Member, IEEE,* Andre K. Geim, Konstantin Novoselov, Frederik Schedin and Peter Blake.

*Abstract*—Graphene - a single atomic layer of graphite – is a recently-found two-dimensional form of carbon, which exhibits high crystal quality and ballistic electron transport at room temperature. Soft magnetic NiFe electrodes have been used to inject polarized spins into graphene and a 10% change in resistance has been observed as the electrodes switch from the parallel to the antiparallel state. This coupled with the fact that a field effect electrode can modulate the conductivity of these graphene films makes them exciting potential candidates for spin electronic devices.

## I. Introduction

Graphene is a name given to an atomic layer of carbon atoms densely packed into a benzene-ring structure with a nearest-neighbor distance of ~1.4Å. This material is widely used in the description of the crystal structure and properties of graphite, large fullerenes and carbon nanotubes. As a first approximation, graphite is made of graphene layers relatively loosely stacked on top of each other with a fairly large interlayer distance of ~3.4Å [1,2]. Carbon nanotubes are usually thought of as graphene layers rolled into hollow cylinders.

The properties of the graphene-based materials are renowned and clearly indicate that planar graphene must be a very remarkable material as well [1,2]. It is expected to be stable at temperatures up to 3000ºC and should be extremely flexible but at the same time as hard to tear apart as diamond, similar in fact to carbon nanotubes. The electronic properties of planar graphene are extraordinary too, with room-temperature mobilities reaching ~15,000 [3] and by improving the technological procedures we have recently improved this by a factor of three. At liquid-helium temperatures graphite routinely exhibits carrier mobilities above $10^6$ cm$^2$/Vs [1,2]. This means that charge carriers in graphene layers can move without scattering at relatively large (submicron) distances even at room temperature. This makes graphene films potentially very important from the point of view of microelectronics applications and for fabricating spintronic devices.

We have found a way of controllably making unbound graphene films from several layers in thickness down to a single layer, which have a nearly macroscopic size (visible by the naked eye). With respect to their electronic properties and despite being effectively only a single atomic layer thick, the films exhibit remarkably high quality so that applying gate voltage can form high-mobility two-dimensional electron and hole gases. We have demonstrated a high-mobility field-effect transistor based on such films [3].

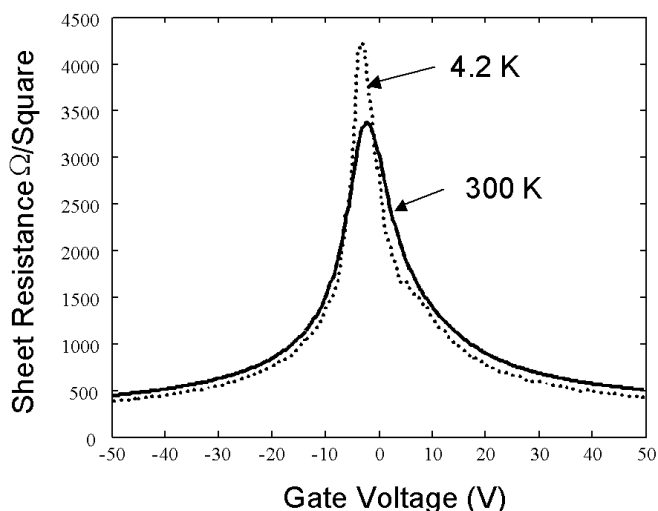

Fig. 1. Sheet Resistance of a graphene flake as a function of gate bias voltage at 4.2 and 300 K. .

Fig. 1 shows the effect of applying a voltage to the silicon substrate which acts as a gate for the field effect induced conductivity in the graphene layer. This gate voltage changes the conductivity by nearly an order of magnitude at 4.2 K and by a factor of 7 at 300 K. Thus if polarized spin injection can be achieved there is significant potential for spin current manipulation in such transistors. Indeed, such spin current manipulation has already been demonstrated in carbon nanotubes [4].

Manuscript received March 13, 2006. This work was supported in part by the EPSRC.

E. W. Hill is with The University of Manchester, School of Computer Science, Oxford Road, Manchester, M13 9PL, UK (phone: +44 (0)161 4552; fax: +44 (0)161 4527; e-mail: e.w.hill@Manchester.ac.uk )

A.K. Geim, is with The University of Manchester, School of Physics and Astronomy, Oxford Road, Manchester, M13 9PL, UK (e-mail: geim@manchester.ac.uk. )

K. Nonoselov, is with The University of Manchester, School of Physics and Astronomy, Oxford Road, Manchester, M13 9PL, UK (e-mail: Konstantin.Novoselov@manchester.ac.uk . )

F. Schedin and P. Blake  are with The University of Manchester, School of Computer Science, Oxford Road, Manchester, M13 9PL, UK

## II. Device Fabrication

Graphene films are made by repeated peeling of small (mm-sized) mesas of highly-oriented pyrolytic graphite (HOPG)[6].

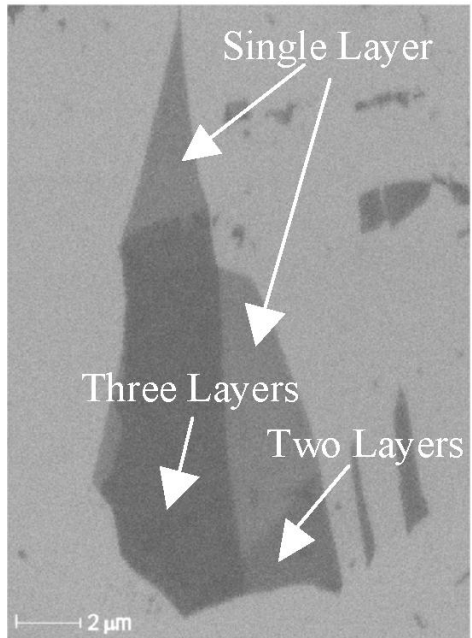

Fig. 2. SEM image of a graphene flake on an $SiO_2$ coated Si substrate. Two single layer regions are shown

The exfoliation continues until we obtain flakes that are nearly invisible in an optical microscope. Fig. 2 shows an SEM image of a graphene flake obtained by this technique and placed on top of an oxidized Si wafer. The thickness of the individual regions is usually confirmed using an AFM and is consistent with the indicated numbers of layers in the graphene stack indicated in fig. 2. One might find it useful to view the obtained films as if they were effectively made from individual carbon nanotubes that are unfolded and stitched together.

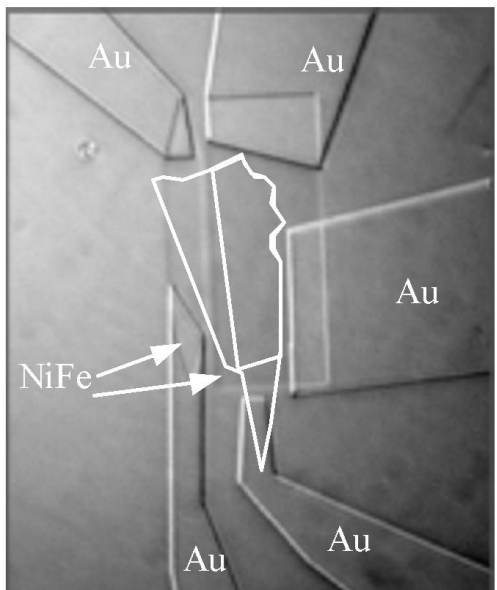

Fig. 3. Differential Phase Contrast (DPC) optical micrograph of a four probe spin valve structure showing the two shaped ferromagnetic contacts and the gold contacts. The outline of the graphene flake has been overlayed on the optical image for clarity.

Spin valve devices are fabricated by defining ferromagnetic NiFe electrodes with a lift off technique. The patterns are produced using electron beam lithography. The flakes are prepared on n-doped silicon wafers with a 300 nm thick $SiO_2$ film on the surface. At this thickness single layer flakes of graphene are just visible in an optical microscope. The silicon substrates have gold crosses defined by conventional optical lithography to allow the flakes to be located and aligned in the electron beam lithography tool.

### 1) Magnetic Electrodes

The permalloy contacts were all 30 nm thick and were deposited onto the patterned resist using electron beam evaporation. The base pressure in the deposition system was $5 \times 10^{-8}$ Torr and a liquid nitrogen cooled cryo-shield was used to reduce water vapour contamination to a minimum during deposition. A forming field of 300 Oe was present on the substrate during the deposition. Lift off was performed using acetone with mild ultrasonic agitation.

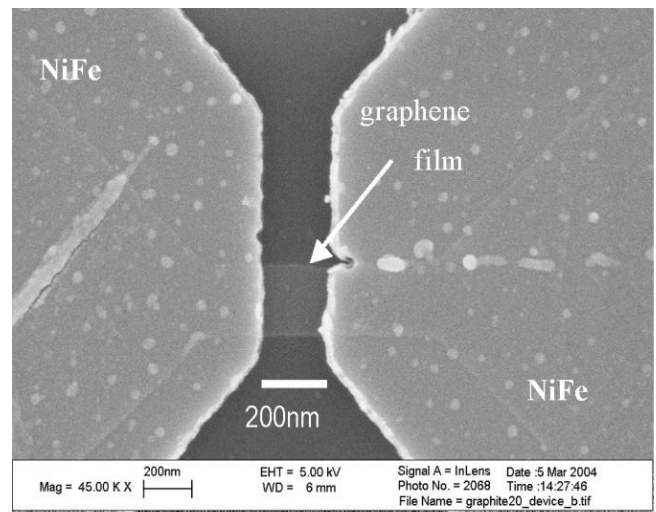

Fig. 4. Two Electrode Spin Valve structure showing NiFe electrodes and graphene wire 200 nm wide.

### 2) Gold Contacts and Bonding.

To make electrical contact to the shaped ferromagnetic electrodes and to allow bonding to a non-magnetic chip carrier a set of gold contacts is deposited. These are patterned using electron beam lithography in a similar manner to the ferromagnetic contacts. A 5nm chromium adhesion layer is used under the gold to improve the contact between the gold and the permalloy. The device is bonded into a non-magnetic chip carrier and the bond wires secured between the contacts on the carrier and the chip using silver loaded epoxy.

Fig. 4 shows a basic two electrode device where the graphene layer has been patterned using electron beam lithography and oxygen plasma etching. The electrodes are fabricated as described above and we rely on imperfections such as the small notch visible in the figure to provide a variation in switching field between the electrodes.

### III. MEASUREMENTS AND RESULTS.

The chip carrier is placed inside a temperature controlled insert. The temperature can be easily varied from 300 K to 77

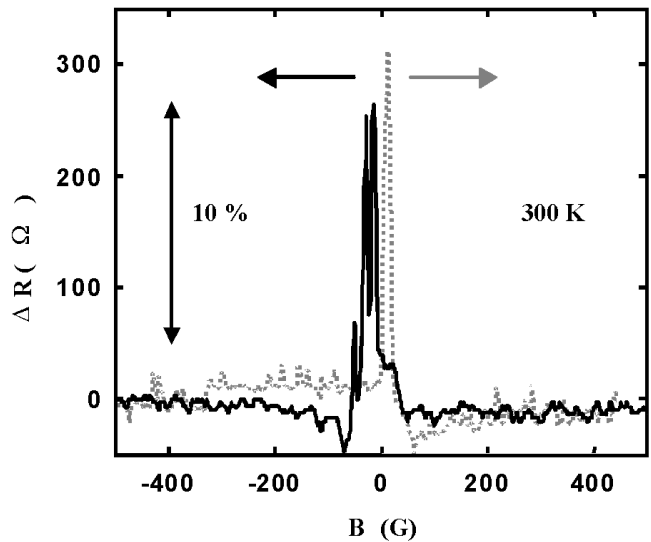

Fig. 5. The magnetoresistive response from a simple two electrode spin valve device with a 200nm wide graphene wire between the electrodes ( as shown in fig 4).

K. A set of coils provides an applied field of up to ±500 Oe. Both two and four terminal devices have been fabricated. In the latter case two gold contacts are made to each ferromagnetic electrode as shown in fig. 3 to allow separate current and voltage connections to be made to the sample.

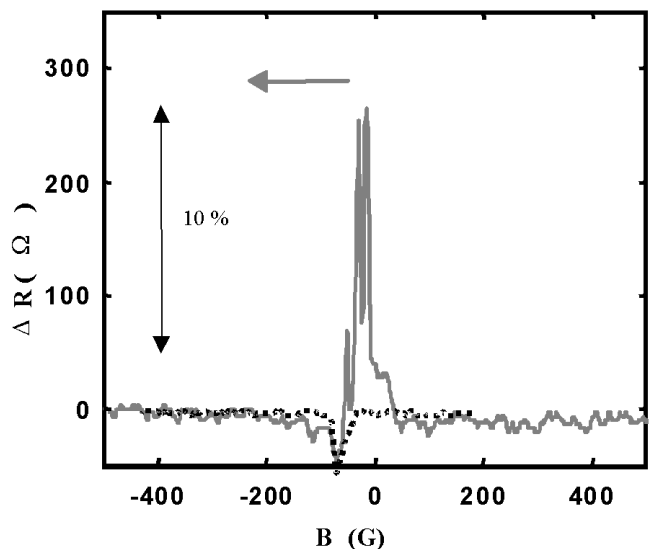

Fig. 6. The magnetoresistive response from a simple two electrode spin valve device (solid curve) with the electrode AMR response from another similar device superimposed (broken curve).

Resistance measurements are made using a Stanford Research SR830 DSP lock-in amplifier to measure the voltage across the device. The ac signal current is typically set at 1 μA pk-pk and the resistance recorded as a function of applied field with data-capture software developed using LabView.

The results shown in fig. 5 are for the simple two-electrode device illustrated in fig. 4. Despite this highly non-optimised structure we observed a 10% change in resistance at 300K as the applied field is swept between +450 G and – 450 G. The 10% change in resistance is much larger than can be attributed to MR effects in the individual electrodes (also measured separately in dedicated experiments), giving confidence that it is due to the spin valve effect with graphene acting as the non-magnetic conductor. Although spin valve effects have been observed in carbon nanotubes [4, 5] this is the first observation of this in planar graphene.

The negative ΔR is mainly due to the Anisotropic Magnetoresistance (AMR) of the electrodes. This is seen in fig. 6 where the AMR response has been directly measured from the electrode of another, similar device, and is superimposed on the results for the negative going field sweep of the graphene device. The sharp peaks can thus be interpreted as the switching at the spin injection edge of the electrode, the bulk of the electrode switching later as indicated by the AMR response. The spikes in the response are due to magnetic domains in the electrodes being sampled by the narrower graphene conductor.

### IV. CONCLUSIONS

We have fabricated a simple spin valve structure using graphene to provide the spin transport medium between the ferromagnetic electrodes. The magnetoresistive response indicated a change which was larger than could be accounted for by AMR effects in the electrodes and so suggests that polarized spin injection and transport in the graphene layer is taking place. We are now working on the more optimized spin valve structures with true four electrode geometry (fig. 3) using both magnetic and non-magnetic contacts to exclude the possibility of interfacial effects between the electrodes and the graphene, such as localized spin channeling, giving rise to the magnetoresistive response reported here.